\documentclass[prc,twocolumn,floatfix,showpacs,preprintnumbers,amsmath,amssymb]{revtex4}

\usepackage{graphicx}
\usepackage{bm}

\def \lf {\left}
\def \ri {\right}

\begin{document}


\title{Neutron skin thickness in droplet model 
with surface width dependence:
\\
indications of softness of the nuclear symmetry energy}

\author{M. Warda\textsuperscript{1,2}}
 \email{warda@kft.umcs.lublin.pl}
\author{X. Vi\~nas\textsuperscript{1}}
 \email{xavier@ecm.ub.es}
\author{X. Roca-Maza\textsuperscript{1}}
 \email{roca@ecm.ub.es}
\author{M. Centelles\textsuperscript{1}}
 \email{mariocentelles@ub.edu}

\affiliation{\textsuperscript{1}Departament d'Estructura i Constituents de la Mat\`eria
and Institut de Ci\`encies del Cosmos, 
Facultat de F\'{\i}sica, Universitat de Barcelona,
Diagonal {\sl 647}, {\sl 08028} Barcelona, Spain\\
\textsuperscript{2}Katedra Fizyki Teoretycznej, Uniwersytet Marii Curie--Sk\l odowskiej,
        ul. Radziszewskiego 10, 20-031 Lublin, Poland}

\date{\today}

\begin{abstract} 
We analyze the neutron skin thickness in finite nuclei with the
droplet model and effective nuclear interactions. The ratio of the
bulk symmetry energy $J$ to the so-called surface stiffness
coefficient $Q$ has in the droplet model a prominent role in driving
the size of neutron skins. We present a correlation between the density
derivative of the nuclear symmetry energy at saturation and the $J/Q$
ratio. We emphasize the role of the surface widths of the neutron and
proton density profiles in the calculation of the neutron skin
thickness when one uses realistic mean-field effective interactions.
Next, taking as experimental baseline the neutron skin sizes measured
in 26 antiprotonic atoms along the mass table, we explore constraints
arising from neutron skins on the value of the $J/Q$ ratio. The
results favor a relatively soft symmetry energy at subsaturation
densities. Our predictions are compared with the recent constraints
derived from other experimental observables. Though the various
extractions predict different ranges of values, one finds a narrow
window $L\sim 45$--75~MeV for the coefficient $L$ that characterizes
the density derivative of the symmetry energy which is compatible with
all the different empirical indications.

\end{abstract}

\maketitle

\section{Introduction}


Neutron skin thickness is the name in common usage to refer to the
difference between the root-mean-square (rms) radii of the neutron and
proton density distributions of atomic nuclei:
\begin{equation} 
\label{skin}
\Delta R_{np}= \langle r^2 \rangle_n^{1/2} 
- \langle r^2 \rangle_p^{1/2} .
\end{equation} 
Experimentally, the value of the proton rms radius $\langle r^2
\rangle_p^{1/2}$ is obtained from the charge radius. The latter has
been measured by electron-nucleus elastic scattering with high
accuracy (often the accuracy in charge radii is better than 1\%
\cite{ang04}). In contrast, our knowledge of the neutron distribution
in nuclei and of its rms radius $\langle r^2 \rangle_n^{1/2}$, as well
as our knowledge of $\Delta R_{np}$, is till date less precise. This
situation looks inadequate in anticipation of the next generation of
rare ion accelerator facilities that are planned for the synthesis and
study of exotic nuclei, as in RIKEN centers (Japan), in FAIR at GSI
(Germany), in the CSR at HIRFL (China), or in FRIB at MSU (USA).
Without securing sufficient knowledge of the neutron distribution in
stable nuclei, the prospects of nuclear structure theory in this
thriving domain may be compromised. It is expected that
parity-violating electron scattering will provide in the nearby future
a leap forward in the quest for high precision determinations of the
neutron radius in heavy nuclei \cite{prex}.

The calibration of the neutron skin thickness of nuclei also is one of
the problems at the forefront of nuclear structure by reason of being
intimately correlated with the nuclear symmetry energy. Indeed, the
symmetry energy is a fundamental quantity in nuclear physics and
astrophysics, because it governs at the same time important properties
of very small entities like atomic nuclei and of very large objects
like neutron stars \cite{lat04}. One of the crucial properties of the
symmetry energy, which still is not sufficiently well constrained, is
its dependence on the nuclear density. It is relevant in the
astrophysical context in order to understand a wealth of phenomena
\cite{lat04,ste05a,hei00}, including supernova explosions, neutrino
emission, and the cooling mechanism of protoneutron stars, as well as
mass-radius relations in neutron stars. Moreover, the density content
of the symmetry energy eventually relates to basic issues in physics.
This is the case of precision tests of the Standard Model through
atomic parity non-conservation observables \cite{sil05}, and even of
studies on constraining a possible time variation of the gravitational
constant \cite{gravi}. In terrestrial laboratories, the available
tools to delineate the density dependence of the symmetry energy at
saturation and subsaturation densities, include the interaction
potential between neutron-rich nuclei \cite{khoa96}, observables like
isospin diffusion and isoscaling in heavy-ion reactions at
intermediate energies \cite{tsa04,che05,li05b,ste05,lie07,li08,
sou06,she07,she07a,fam06,gai04,bar05,tsa09}, different modes of
collective excitations of nuclei \cite{pie02,li07,kli07,tri08,lia08},
and, of course, data on the binding and structure of neutron-rich
nuclei and on their neutron skin thickness.

In the literature there exist several theoretical formulations to
investigate the neutron skin thickness of neutron-rich nuclei and its
connections with the symmetry energy. This is the case, for instance, of
methods based on the droplet model \cite{mye80,swi05}, on the concept
of surface symmetry energy \cite{ste05a,dan03,dan09}, thermodynamical
arguments \cite{pet96}, nucleonic density form factors \cite{miz00},
mean-field analyses \cite{tre86,fur02,war98}, or studies in the spirit
of the Landau-Migdal approximation \cite{die03}. It has been shown
that the neutron skin thickness in heavy nuclei, like $^{208}$Pb,
calculated in mean-field models with either non-relativistic or
relativistic effective nuclear interactions, displays a linear
correlation with the slope of the neutron equation of state (EOS)
obtained with the same interactions at a neutron density $\rho \approx
0.10$ fm$^{-3}$ \cite{bro00,typ01,cen02}. A similar correlation exists
between $\Delta R_{np}$ and the density derivative of the bulk
symmetry energy \cite{che05,li05b,lie07,li08,fur02,ava07,bal04}, as
the latter is a measure of the pressure difference between neutrons
and protons. These correlations have been exploited in recent years to
gain a deeper understanding of the isospin properties of the effective
nuclear interaction and to relate them with nuclear and astrophysical
observations. 

The rms radius of neutron densities in nuclei has been measured with
hadronic probes such as proton-nucleus elastic scattering
\cite{ray78,sta94,kar02,cla03} or inelastic scattering excitation of the
giant dipole and spin-dipole resonances \cite{kra99,kra04}. On the
other hand, antiprotonic atoms are helpful to probe the size of the
neutron skin of nuclei from the fact that the nuclear periphery is
very sensitive to antiprotons in the normally electronic shell.
Experimentalists combine two different techniques in this case
\cite{trz01,jas04,klo07}, namely, the measurement of the antiprotonic
X-rays which determine the atomic level shifts and widths due to the
strong interaction, and the radiochemical analysis of the yields after
the antiproton annihilation. The values of the neutron skin thickness
of 26 stable nuclei from $^{40}$Ca to $^{238}$U deduced from
antiprotonic atoms data by Trzci{\'n}ska et al.\ \cite{trz01,jas04}
follow a roughly linear trend with the overall relative neutron excess
$I=(N-Z)/A$ of these nuclei. This trend can be fitted by the
relationship $\Delta R_{np}= (0.90 \pm 0.15) I + (-0.03 \pm 0.02)$ fm
as discussed in Refs.\ \cite{trz01,jas04}. As mentioned, all neutron
skin thickness measurements have relatively large uncertainties in
comparison with charge radii, and sometimes the results from different
experimental techniques are not totally consistent among them
\cite{kra04,jas04}. The neutron skin sizes determined in Refs.\
\cite{trz01,jas04} from the analysis of antiprotonic atoms are till
date the largest set of {\it uniformly} measured values of $\Delta
R_{np}$ {\it all over} the periodic table ($40 \leq A \leq 238$). Due
to this reason, we shall use hereinafter these data as the
experimental benchmark for our calculations.

The droplet model (DM) describes in a physically transparent way
nuclear radii and relates them directly with basic properties of the
nuclear interactions. In the present paper we study the neutron skin
thickness of atomic nuclei with the DM using various effective nuclear
interactions of the Skyrme, Gogny, and relativistic mean-field (RMF)
type. The present work extends with a new analysis and perspective a
first presentation of our study made in Ref.\ \cite{cen09}. Here, we
will show that the ratio of the DM parameters $J$ and $Q$, which
drives the value of the neutron skin thickness in heavy nuclei, is
correlated with the slopes in density of the nuclear symmetry energy
and of the EOS of neutron matter. We compare the DM values for the
neutron skin thickness with the results obtained in self-consistent
extended Thomas-Fermi (ETF) calculations of finite nuclei
\cite{bra85,cen90,cen93,cen98a}, since both methods are free of shell
effects. A non-negligible role of the contribution of the difference
in the surface widths of the neutron and proton density profiles is
noticed. Next, we use the experimental neutron skin thickness measured
in antiprotonic atoms to explore the range of possible values of the
ratio $J/Q$ that are favored by neutron skins. With these values we
can predict some properties of the density dependence of the nuclear
symmetry energy. Our results are compared with the constraints
recently obtained in the literature using other observables and
methods.

The present paper is arranged as follows. In the second section we
study the neutron skin thickness of heavy nuclei on the basis of the
DM \cite{mye69,mye77,mye80} and show a correlation that links the
value of the ratio $J/Q$, which governs the neutron skin thickness of
nuclei, with the slope of the symmetry energy in bulk matter at
saturation. In the third section, the contribution of the surface
widths of the neutron and proton density distributions to the neutron
skin thickness is analyzed with the DM using non-relativistic and
covariant mean-field nuclear interactions. In the fourth section we
estimate possible constraints on the density dependence of the nuclear
symmetry energy on the basis of the DM and the experimental data on
neutron skin sizes derived from antiprotonic atoms. We discuss the
present results in comparison with the recent constraints obtained from
various observables and methods. Finally, the summary and our
conclusions are laid in the fifth section. We outline the procedure
for the calculation of the $Q$ coefficient in the Appendix.

\section{The framework}


\subsection{Neutron skin thickness in the droplet model}

In the DM of average nuclear properties \cite{mye69,mye77,bra85} the
neutron skin thickness of a finite nucleus is computed from the
expression \cite{mye69,mye80}
\begin{equation}
\label{skincoulDM}
\Delta R_{np}=\sqrt{\frac{3}{5}}\left[t - \frac{e^2 Z}{70J}
+\frac{5}{2R}\left(b_n^2-b_p^2\right) \right] ,
\end{equation}
where $e^2 Z/70J$ is a correction due to the Coulomb interaction,
$R=r_0 A^{1/3}$ is the nuclear radius, and $b_n$ and $b_p$ are the
surface widths of the neutron and proton density profiles. In the
``standard'' version of the DM it is assumed that $b_n=b_p=1$~fm
\cite{mye69,mye80,swi05}, which implies a vanishing surface width
correction to the neutron skin thickness.

The quantity $t$ in Eq.\ (\ref{skincoulDM}) represents the distance
between the neutron and proton mean surface locations. This distance
is computed as \cite{mye69,mye80}
\begin{equation}
\label{tDM}
t=\frac{3}{2} r_0 \, \frac{J}{Q}\;
\frac{\displaystyle I-\frac{c_1 Z}{12J}A^{-1/3} }
{\displaystyle 1+\frac 94 \frac JQ A^{-1/3} } \;,
\end{equation}
where $I=(N-Z)/A$, $J$ is the symmetry energy coefficient at
saturation, $Q$ is the surface stiffness coefficient, and $c_1 = 3e^2
/5r_0$. The coefficient $J$ represents with a very good accuracy the
energy cost per nucleon to convert all protons into neutrons in
symmetric infinite nuclear matter at saturation density $\rho_0$. The
surface stiffness coefficient $Q$ measures the resistance of the
system against separation of neutrons from protons to form a neutron
skin. To extract $Q$ from an effective nuclear interaction requires to
perform calculations of asymmetric semi-infinite nuclear matter
(ASINM). Therefore, the calculated value of $Q$ may depend somewhat on
the type of approach, such as the Hartree-Fock or Thomas-Fermi
methods, employed to describe the nuclear surface
\cite{far78,bra85,kol85,cen93a,cen98a,dan09}.

\begin{figure}
\includegraphics[width=0.96\columnwidth, clip=true]{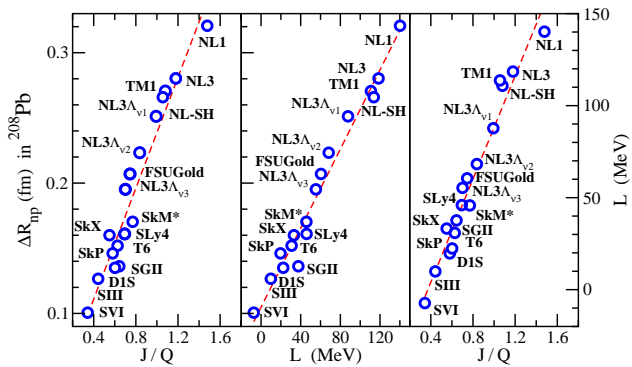}
\caption{(Color online) 
For several nuclear mean-field models, existing correlation between
the neutron skin thickness $\Delta R_{np}$ in $^{208}$Pb and the ratio
$J/Q$ (left panel), and between $\Delta R_{np}$ in $^{208}$Pb and the
slope of the symmetry energy $L$ (middle panel). The correlation
between the coefficient $L$ and the ratio $J/Q$ is also shown (right
panel). In the present figure, $\Delta R_{np}$ has been computed with
Eq.\ (\ref{skin}) from the rms radii of quantal self-consistent
calculations for the indicated mean-field models.
\label{nsfig1}}
\end{figure}

From Eq.\ (\ref{tDM}), one sees that the leading contribution to $t$
in large nuclei is the term $\frac{3}{2} r_0 (J/Q) I$. Thus, the DM
suggests that one can expect a correlation between $\Delta R_{np}$ and
$J/Q$ in heavy nuclei. We illustrate this fact in the left panel of
Fig.~1. We depict there the neutron skin thickness of $^{208}$Pb
obtained from {\em self-consistent quantal}\/ calculations with the
Skyrme and Gogny Hartree-Fock methods as well as with the RMF Hartree
approach. The results are shown as a function of the value of the
$J/Q$ ratio for various mean-field effective interactions. The $J$
values of the selected interactions (about 27--32~MeV in the
non-relativistic forces and about 32--45~MeV in the covariant forces,
see Table~I) cover widely the plausible physical range of the bulk
symmetry energy. The values of $Q$ used in this work have been
extracted from ASINM calculations performed in the extended
Thomas-Fermi (ETF) approach as described in the Appendix (see also
Ref.\ \cite{cen98a}). Even if the shell effects, present in the
mean-field calculations of $\Delta R_{np}$, are not built in the DM
\cite{mye80}, one observes a considerably linear correlation between
the values of $\Delta R_{np}$ and $J/Q$. It should be pointed out that
while all the effective interactions have been accurately calibrated
to data on binding energies and charge radii, and describe these
properties very successfully, they predict widely different values of
$\Delta R_{np}$, as we see in the present case of $^{208}$Pb. This
underlines the fact that the isospin sector of the effective
interactions is little constrained.

\begin{table}[t]
\caption{\label{Table1} Saturation density $\rho_0$ and $J$, $L$,
$K_{sym}$ and $Q$ parameters of the Skyrme and Gogny forces as well as
RMF parameter sets used in this work. }
\begin{center}
\begin{tabular}{lccrrcc}
\hline
force& $\rho_0$  & $J$ & $L\;\;\;$ & $K_{sym}$&$Q$ &$J/Q$\\
     & fm$^{-3}$ & MeV & MeV       & MeV      &MeV & \\ 
\hline
SGII			&0.158&26.83&37.7	&$-$146&41.7&0.64\\
SVI			&0.144&26.88&$-$7.3	&$-$471&78.4&0.34\\
SIII			&0.145&28.16&9.9	&$-$394&63.6&0.44\\
T6			&0.161&29.97&30.9	&$-$211&47.8&0.63\\
SkP			&0.163&30.00&19.7	&$-$267&52.1&0.58\\
SkM*			&0.160&30.03&45.8	&$-$156&39.0&0.77\\
SkX			&0.155&31.10&33.2 	&$-$252&56.2&0.55\\
NL3 $\Lambda_v=0.03$	&0.148&31.68&55.3	&$-8$&45.2&0.70\\
D1S			&0.163&31.93& 22.4 	&$-$252&53\footnotemark[1]$\;$&0.60\\
SLy4			&0.160&32.00&46.0	&$-$120&46.1&0.69\\
FSUGold			&0.148&32.59&60.5	&$-$52&43.7&0.75\\
NL3 $\Lambda_v=0.02$	&0.148&33.15&68.2	&$-$54&39.6&0.84\\
NL3 $\Lambda_v=0.01$	&0.148&34.96&87.7	&$-$46&35.2&0.99\\
NL-SH			&0.146&36.12&113.7	&80&34.5&1.05\\
TM1			&0.145&36.89&110.8	&34&34.3&1.08\\
NL3			&0.148&37.40&118.5	&101&31.7&1.18\\
NL1			&0.152&43.46&140.2	&143&29.4&1.48\\
NL2			&0.146&45.12&133.4	&20&41.7&1.08\\
\hline
\end{tabular}
\end{center}
\footnotetext[1]{Estimated value.}
\end{table}

\subsection{Properties of the nuclear symmetry energy}

Let us consider the energy per particle $e(\rho,\delta)$ in
asymmetric infinite nuclear matter of total density
$\rho=\rho_n+\rho_p$ and relative neutron excess
$\delta=(\rho_n-\rho_p)/\rho$, where $\rho_n$ and $\rho_p$ stand for
the neutron and proton densities, respectively. The general expression
\begin{equation}
 e(\rho,\delta) = 
 e(\rho,\delta=0) + c_{sym}(\rho)\delta^2 + {\cal O}(\delta^4) 
\label{easym}
\end{equation}
defines the symmetry energy coefficient $c_{sym}(\rho)$ of a nuclear
EOS at the density $\rho$. This expression is particularly useful
because $c_{sym}(\rho)$ dominates the corrections to the symmetric
limit for all values of $\delta$, especially at the subsaturation
densities of relevance for finite nuclei \cite{piek09}. Actually,
$c_{sym}(\rho)$ provides with excellent accuracy the difference
between the binding energies of pure neutron matter ($\delta=1$) and
symmetric matter ($\delta=0$).

It is customary, and insightful, to characterize the behavior of an
EOS around the saturation density $\rho_0$ by means of a few bulk
parameters calculated at the saturation point, as in the formula
\cite{bar05,che05,li05b,lie07,li08,mye69,lop88,piek09,bednarek09}
\begin{equation}
\label{erhodel}
e(\rho,\delta) \approx
a_v + \frac{K_v}{2} \epsilon^2+ \left[ J- L\epsilon
+\frac{K_{sym}}2 \epsilon^2\right]\delta^2 ,
\end{equation}
where $\epsilon=(\rho_0-\rho)/3\rho_0$ expresses the relative density
displacement from $\rho_0$. Here, the quantities $a_v$ and $K_v$
denote the energy per particle and the incompressibility modulus of
symmetric nuclear matter. One has $c_{sym}(\rho_0)= J$. The DM
coefficients $L$ and $K_{sym}$ are, respectively, proportional to the
slope and the curvature of the symmetry energy coefficient
$c_{sym}(\rho)$ at saturation density:
\begin{equation}
L = \lf. 3\rho_0\frac{\partial c_{sym}(\rho)}{\partial \rho}
\ri|_{\rho_0} , \quad
K_{sym} = \lf. 9\rho_0^2\frac{\partial^2
c_{sym}(\rho)}{\partial\rho^2} \ri|_{\rho_0} .
\label{l_ksym}
\end{equation}
The quadratic expansion $c_{sym}(\rho) \approx J- L\epsilon
+\frac{1}{2}K_{sym} \epsilon^2$ in Eq.\ (\ref{erhodel}) is often a
reliable representation of the actual value of the $c_{sym}(\rho)$
coefficient at densities roughly between $\rho_0/2$ and $2\rho_0$
\cite{piek09}. For instance, in the case of a typical subsaturation
density value $\rho= 0.10$ fm$^{-3}$, one finds that the above
quadratic expansion of $c_{sym}(\rho)$ differs from the exact
$c_{sym}(\rho)$ by less than 1\% in many different nuclear mean field
forces \cite{cen09}. These facts point to the usefulness of
investigating parameters such as $L$ and $K_{sym}$ for the
characterization of the density dependence of the symmetry energy.

The values of the DM coefficients $J$, $L$, and $K_{sym}$ for the
non-relativistic forces and the RMF parameter sets considered in this
work are given in Table~I. In the Skyrme and Gogny effective
interactions the symmetry energy coefficient at saturation $J$ takes
values around 30 MeV\@. The RMF parameterizations have larger values
of $J$, also with a larger spread. The slope ($L$) and the curvature
($K_{sym}$) of the symmetry energy at saturation take even more widely
scattered values among the different interactions. The consequence is
that all mentioned nuclear models predict a different behavior of the
symmetry energy at subsaturation densities, what can be seen e.g.\ in
Fig.~1 of Ref. \cite{che05}.

As is known, the density dependence of the symmetry energy near
saturation tends to be much softer in the non-relativistic forces than
in the covariant meson-exchange models of nuclear structure (see the
values of $L$ in Table~I). In particular, taking into account Eq.\
(\ref{erhodel}), the slope of the neutron EOS
\begin{equation}
\frac{d e(\rho,\delta=1)}{d \rho} =
\frac{L}{3 \rho_0} - \frac{K_v+K_{sym}}{3 \rho_0} \epsilon \,,
\label{slopeneu}
\end{equation}
and of the symmetry energy
\begin{equation}
\frac{d c_{sym}(\rho)}{d \rho} =
\frac{L}{3 \rho_0} - \frac{K_{sym}}{3 \rho_0} \epsilon \,,
\label{slopesym}
\end{equation}
calculated at densities $\rho$ close to the saturation value $\rho_0$
differ considerably between models. We
also know that $\Delta R_{np}$ in $^{208}$Pb shows a linear dependence
with these slopes at some subsaturation density $\rho \simeq 0.10$
fm$^{-3}$ \cite{bro00,typ01,cen02,bal04}. Therefore it is reasonable
that the neutron skin thickness in $^{208}$Pb and the leading term $L/3
\rho_0$ of Eqs. (\ref{slopeneu}) and (\ref{slopesym}) are related. As
far as $\rho_0$ does not change much in the different effective forces,
a correlation between $\Delta R_{np}$ in $^{208}$Pb and the DM
coefficient $L$ is thus expected \cite{che05,li05b,lie07,fur02,ava07}
and we display it in the middle panel of Fig.~1.

From the discussed results of $\Delta R_{np}$ versus $J/Q$ and of
$\Delta R_{np}$ versus $L$ in Fig.~1, a correlation between the DM
coefficient $L$ and the $J/Q$ ratio is to be expected too. Note that
$L$ rules the density dependence of the symmetry energy of the nuclear
equation of state [Eq.\ (\ref{slopesym})], and that $Q$ governs the
thickness of the neutron skin of finite nuclei [Eqs.\
(\ref{skincoulDM}) and (\ref{tDM})]. The correlation between $L$ and
$J/Q$ can be seen in the right panel of Fig.~1. This correlation shows
that the value of the slope $L$ of the symmetry energy at the
saturation density increases with the value of the ratio between the
bulk symmetry energy coefficient $J$ and the surface stiffness
coefficient $Q$. The trend is considerably linear among the various
nuclear effective interactions.

Previous literature \cite{kol85,cen98a} has shown that the systematics
of experimental binding energies relates increasing values of $J$ with
decreasing values of $Q$ in nuclear effective interactions whose
parameters have been adjusted to describe experimental data. We note
this same trend in Table~I, where the RMF sets that in general have
larger $J$ values also tend to have smaller $Q$ values than their
non-relativistic counterparts. A smaller $Q$ coefficient means that it
is easier to develop a neutron skin in finite nuclei. Consistently,
the neutron skin thickness in $^{208}$Pb (or any other heavy nucleus)
is usually larger when computed with a RMF parameter set than when
computed with a non-relativistic force. These facts, and the noticed
correlation between $L$ and $J/Q$, allow one to interpret in a
qualitative way within the DM, the correlation pointed out in Refs.\
\cite{bro00,typ01,cen02,bal04} between the slope of the symmetry
energy at some subsaturation density $\rho \simeq 0.10$ fm$^{-3}$ and
the neutron skin thickness in a heavy nucleus. In a recent work
\cite{cen09} we have investigated further the relations of the neutron
skin thickness with the parameters $L$ and $K_{sym}$ that characterize
the density dependence of the symmetry energy around saturation.


\section{Surface width contribution to the neutron skin thickness}


The neutron skin thickness values derived from measurements performed
in antiprotonic atoms have been obtained in Refs.\ \cite{trz01,jas04}.
It is assumed that the neutron skin is due to an enhancement of the
neutron surface width with respect to the proton surface width, and
that the mean location of the proton and neutron surfaces in these
nuclei are the same. This situation corresponds to the so-called
``neutron halo-type'' distribution \cite{trz01}. It has been shown
that the same set of experimental values of neutron skin thickness can
be explained with similar quality as in \cite{trz01} by means of the
``standard'' version of the DM (where $b_n=b_p$) \cite{swi05}. The
latter case assumes that the peripheral neutrons are concentrated at
the neutron surface, which is shifted with respect to the proton
surface, and that both the neutron and proton density distributions
have the same surface width. This is rather the pattern of the
so-called ``neutron skin-type'' distribution according to Ref.\
\cite{trz01}.

The analysis of neutron and proton densities calculated with nuclear
mean field interactions carried out in Ref.\ \cite{miz00} by means of
the Helm model, points out that self-consistent mean field densities
show a mixed character between the ``neutron halo'' and ``neutron
skin'' patterns. This means that, actually, the self-consistent
neutron and proton density profiles obtained with nuclear effective
interactions differ not only in the mean location of their surfaces
but also in their surface widths. In the following we shall see that
similar conclusions are found from the calculations of the neutron
skin thickness performed in the DM with formula (\ref{skincoulDM}). It
will turn out that the surface width contribution $\propto
(b_n^2-b_p^2)$ in the DM expression (\ref{skincoulDM}) for the neutron
skin thickness, which arises from $b_n \neq b_p$, is necessary to
reproduce the neutron skin thickness values calculated from the
definition (\ref{skin}) using self-consistent densities of finite
nuclei obtained with the ETF approach in mean-field theory, for
non-relativistic forces as well as for relativistic parameterizations.

In Fig.~2 we display by empty symbols, as a function of the overall
relative neutron excess $I=(N-Z)/A$, the neutron skin thickness
predicted by the ``standard'' version of the DM (namely, Eq.\
(\ref{skincoulDM}) with $b_n=b_p$) using some well-known effective
forces. The nuclei are those from $^{40}$Ca to $^{238}$U measured in
the experiments with antiprotonic atoms \cite{trz01,jas04}, and which
were studied with the DM in Ref.\ \cite{swi05}. The values shown in
Fig.~2 have been computed using the SIII and SkM* Skyrme forces and
the NL-SH and NL3 RMF parameter sets, as suitable examples. We have
chosen these four parameter sets for display in Fig.~2 because they
span the whole range of values of the ratio $J/Q$ of nuclear
interactions that describe reasonably well the ground-state properties
of finite nuclei, having a bulk symmetry energy coefficient $J$
between 28 MeV and 37 MeV (see Fig.~1 and Table~I). The DM results for
$\Delta R_{np}$ obtained with the other mean field interactions
considered in this work that have a $J$ coefficient between the values
of SIII and NL3, also lie within the window of results delimited by
the SIII and NL3 interactions in Fig.~2.

The values of $\Delta R_{np}$ predicted by the DM are compared in
Fig.~2 with the values that we obtain from self-consistent ETF
calculations in finite nuclei (filled symbols). Both models do not
incorporate shell effects. In Fig.~2 we have used the ETF approach in
the non-relativistic \cite{cen90} and relativistic \cite{cen93}
frameworks to compute $\Delta R_{np}$. We have calculated these ETF
values of $\Delta R_{np}$ by application of Eq.\ (\ref{skin}) with the
rms radii of the self-consistent neutron and proton densities obtained
in each isotope. From Fig.~2 two significant points stem. First, the
predictions of the DM in the ``standard'' form ($b_n=b_p$)
systematically undershoot the ETF neutron skin thickness computed in
finite nuclei with the selected effective nuclear interactions. In
particular, this trend is reinforced with growing neutron excess $I$.
Second, it can be observed that for a given nucleus the difference
between the ETF value of $\Delta R_{np}$ computed with (\ref{skin})
and the value provided by the ``standard'' DM prescription is slightly
larger in the RMF parameter sets than in the Skyrme forces.
Altogether, these facts suggest that in the mean-field interactions
the surface width contribution to the DM formula for $\Delta R_{np}$
does not vanish, and that this contribution has some dependence on the
ratio $J/Q$ of the force.

\begin{figure}
\includegraphics[width=0.96\columnwidth, angle=0, clip=true]{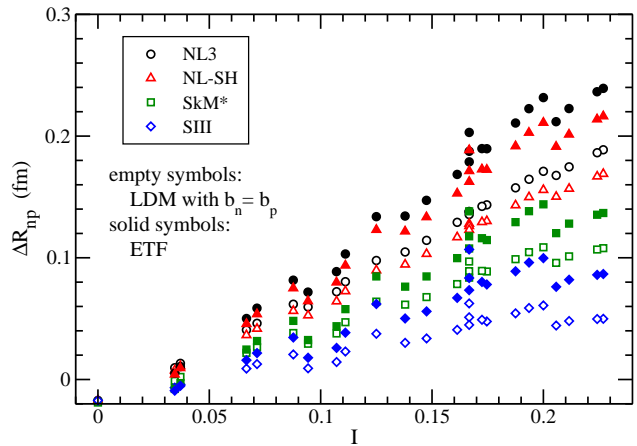}
\caption{(Color online) The neutron skin thickness predicted by the
``standard'' version of the DM (Eq.\ (\ref{skincoulDM}) with
$b_n=b_p$) is compared with the result obtained from self-consistent
ETF calculations of finite nuclei, in four illustrative mean-field
parameter sets. The nuclei considered are those investigated
experimentally with antiprotonic atoms in Refs.~\cite{trz01,jas04} and
have masses $40 \leq A \leq 238$.
\label{nsfig2}}
\end{figure}

To apply the full Eq.\ (\ref{skincoulDM}) to compute the neutron skin
thickness including the surface width correction, one needs to evaluate
the neutron and proton surface widths in finite nuclei. In practice,
there is not conclusive experimental evidence on the difference of the
surface widths $b_n$ and $b_p$ of the nuclear density distributions
\cite{swi05,klo07,mye80,bra85,miz00}. To estimate $b_n^2 - b_p^2$ we
will therefore rely on theoretical guidance as a surrogate. Within the
context related to the DM and the leptodermous expansion of a finite
nucleus \cite{mye69,mye77,bra85}, the surface properties in finite
nuclei can be extracted from ASINM calculations. The semi-infinite
geometry does not include shell, Coulomb or finite-size effects. We
will use here the ETF method including $\hbar^2$ corrections for
describing ASINM, since the ETF method is free of the Friedel
oscillations of the quantal densities \cite{cen93a,dan09}. We
summarize in the Appendix the basic aspects of the procedure.
More details can be found in Ref.\ \cite{cen98a}. From Eqs.\
(\ref{profile})--(\ref{widths}) of the Appendix, we can obtain the
values of the surface widths $b_n$ and $b_p$ in ASINM with a given
relative neutron excess in the bulk $\delta_0$. In the DM, these
surface widths correspond to the values $b_n$ and $b_p$ in finite
nuclei if $\delta_0$ is calculated from the overall relative neutron
excess $I$ of the nucleus through the following relation
\cite{mye69,bra85,cen98a}:
\begin{equation}
\delta_0 = \frac {\displaystyle I+\frac38 \frac{c_1}{Q} \frac{Z^2}{A^{5/3}}}
{\displaystyle 1 + \frac{9}{4} \frac{J}{Q} A^{-1/3}},
\label{delta0}
\end{equation}
which takes into account the Coulomb correction.

Once the neutron and proton surface widths in finite nuclei are known,
we can compute their contribution to the neutron skin thickness, which
reads [cf.\ Eq.\ (\ref{skincoulDM})]
\begin{equation}
\Delta R_{np}^{sw}=\sqrt{\frac{3}{5}}
\frac{5}{2R}\left(b_n^2-b_p^2\right)\,.
\label{swidthns}
\end{equation}
The corresponding values of $\Delta R_{np}^{sw}$ for the nuclei
considered in Fig.~2 are displayed in the bottom panel of Fig.~3. It is
worth to point out that the calculated $\Delta R_{np}^{sw}$ values
show, for each nuclear interaction, a well-defined increasing linear
trend as a function of the overall relative neutron excess $I$ of the
nuclei.

The neutron skin thickness predictions of the DM when one includes the
surface width contribution (\ref{skincoulDM}) are displayed in the top
panel of Fig.~3 by empty symbols. Note that the results correspond to
adding the values $\Delta R_{np}^{sw}$ shown in the bottom panel of
this figure to the DM values that we have displayed in Fig.~2. As done
in Fig.~2, we compare in the top panel of Fig.~3 the DM results with
the self-consistent ETF calculations of $\Delta R_{np}$
\cite{cen90,cen93}. One now observes an improved and remarkable
agreement between the DM predictions and the self-consistent ETF
values computed with the same interaction, stemming from the inclusion
of the calculated $\Delta R_{np}^{sw}$ contribution. It is interesting
to note that the neutron skin thickness obtained with Eq.\
(\ref{skin}) from the rms radii of the self-consistent ETF
calculations in finite nuclei shows a well defined increasing linear
tendency with the relative neutron excess $I$, similarly to the case
of the results of the DM and in consonance with the trend of the
experimental values derived from antiprotonic atoms \cite{trz01}.

\begin{figure}
\includegraphics[width=0.96\columnwidth, angle=0, clip=true]{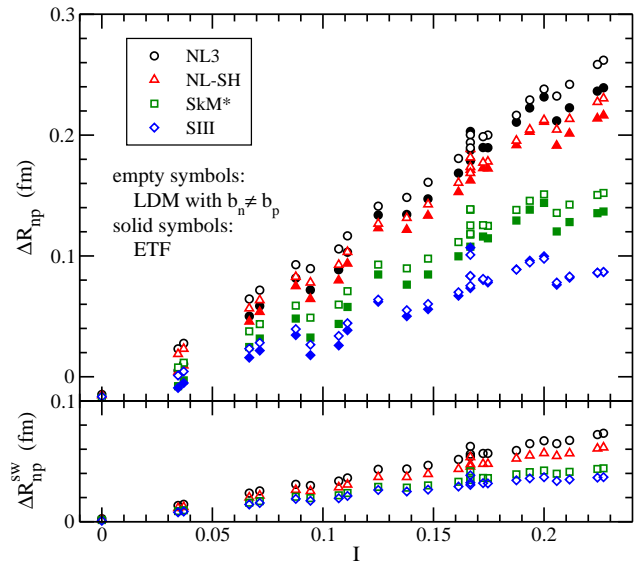}
\caption{(Color online) Upper panel: the same as in
Fig.~(\ref{nsfig2}) but here the DM values include a non-vanishing
surface width contribution $\Delta R_{np}^{sw}$ [Eq.~(\ref{swidthns})]
with $b_n$ and $b_p$ obtained from ASINM calculations as described in
the text. Lower panel: the surface width contribution $\Delta
R_{np}^{sw}$ (the vertical scale proportionality is the same as in the
upper panel).
\label{nsfig3}}
\end{figure}

The lower panel of Fig.~3 also suggests that, for a given nucleus,
$\Delta R_{np}^{sw}$ grows when the $J/Q$ ratio of the nuclear
interaction increases (see Table~I). To analyze this behavior in more
detail, we fit $\Delta R_{np}^{sw}$ by means of a law $\sigma^{sw} I$,
which defines the slope $\sigma^{sw}$ of $\Delta R_{np}^{sw}$ with
respect to the relative neutron excess $I$. This slope is displayed in
Fig.~4 as a function of the $J/Q$ ratio for different interactions.
The slopes $\sigma^{sw}$ lie inside a band limited by two straight
lines, corresponding to the equations $\sigma^{sw} = 0.3 J/Q +0.07$ fm
(left) and $\sigma^{sw} = 0.3 J/Q - 0.05$ fm (right). It is worth to
notice that all considered Skyrme forces have slopes $\sigma^{sw}$
below 0.25~fm, whereas the analyzed RMF models have slopes
$\sigma^{sw}$ always above this value.

\begin{figure}
\includegraphics[width=0.96\columnwidth,  clip=true]{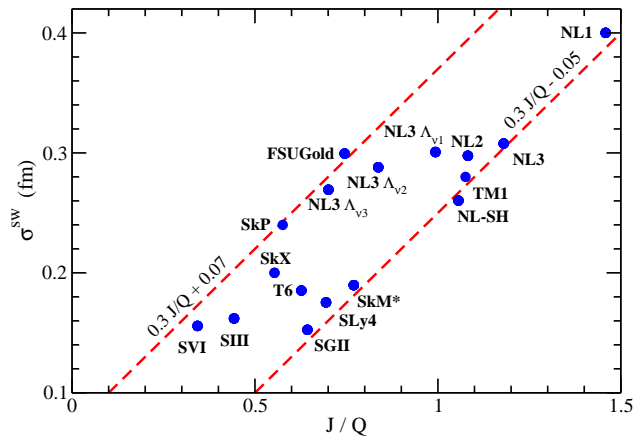}
\caption{(Color online) Average slope of $\Delta R_{np}^{sw}$ with
respect to $I$ for various nuclear mean-field models as a function of
$J/Q$. All the data lie in the area limited by the marked lines
$\sigma^{sw} = 0.3 J/Q + 0.07$ fm and $\sigma^{sw} = 0.3 J/Q - 0.05$.
\label{fig5}}
\end{figure}

One relevant conclusion is that bulk and finite nuclei properties
described through successful theoretical mean-field models constrain
the possible values of the surface width contribution to the neutron
skin thickness between the limits portrayed in Fig.~4. From this figure
it can be deduced that the surface width contribution to the neutron
skin thickness $\Delta R_ {np}^{sw}$ has, on top of a global increasing
trend with $J/Q$, a more involved dependence on the parameters of the
effective nuclear interactions. For instance, on the one hand, the RMF
force FSUGold and the Skyrme force SkM* have almost the same $J/Q$
ratio (see Table~I). However, the predicted values of $\sigma^{sw}$
are clearly different for both interactions as can be seen in Fig 4.
On the other hand, it is possible to find different interactions which
have almost the same slope $\sigma^{sw}$, and therefore the same
$\Delta R_{np}^{sw}$, but with different values of the $J/Q$ ratio.
Some examples of this fact in Fig.~4 are the Skyrme forces SIII and
SGII (with slopes of 0.16 fm and 0.15 fm, respectively), and the RMF
parameterizations FSUGold and NL3 (with slopes 0.30 fm and 0.31 fm).
These facts suggest that it is possible to find the same total neutron
skin thickness by combining a small value of the $J/Q$ ratio in $t$
[see Eq.\ (\ref{tDM})] with a large $\Delta R_{np}^{sw}$ contribution
or, vice versa, by combining a large $J/Q$ ratio in $t$ with a small
$\Delta R_{np}^{sw}$ contribution.

\section{Estimates of the density content of the symmetry energy}


As we have pointed out, the nuclear symmetry energy and the properties
of the EOS of neutron-rich matter are of increasing importance in both
nuclear physics and astrophysics. There is a significant recent effort
in the community towards constraining the values of the parameters
that characterize the density dependence of the symmetry energy in the
subsaturation regime of the EOS\@. The new developments come in
particular from the investigation of isospin-sensitive observables in
intermediate-energy heavy-ion collisions
\cite{tsa04,che05,li05b,ste05,lie07,li08, 
sou06,she07,she07a,fam06,gai04,bar05,tsa09} and in nuclear resonances
\cite{pie02,li07,kli07,tri08}. Obviously, the different studies do not
deal with exactly the same regimes of density and energy. One also has
to keep in mind that the connection of the experiments with the EOS is
not at all trivial; it often requires of extrapolations of the
measured data, which imply a model dependence. Therefore, it is
important to further investigate indications from those and other
experimental probes of the symmetry energy, with different
methodologies, as well as to study the interplay between the
constraints derived from the different analyses.

In the present section we want to apply the experience gained in the
DM study of the neutron skin thickness performed in the previous
sections, to estimate possible constraints on the density dependence
of the nuclear symmetry energy as suggested by neutron skin data. To
this end, we shall first obtain the range of values of the $J/Q$ ratio
which are compatible with the neutron skin thickness derived from the
experimental data in antiprotonic atoms. We recall that this set of
data for 26 stable nuclei is till date the largest set of uniformly
measured neutron skins spanning the mass table. Once the estimated
range of values for the ratio $J/Q$ will be found, the constraints
derived from the neutron skin data on the DM parameter $L$, related to
the density derivative of the symmetry energy, will be determined from
the existing linear correlation between the values of $L$ and $J/Q$.
This correlation has been displayed in Fig.~1 and is a general feature
of mean field interactions that have been adjusted to reproduce with
good accuracy binding energies and charge radii (often among other
properties) of nuclei across the periodic table.

\subsection{Constraints on the {\boldmath$J/Q$} ratio}

From the previous section we know that the surface width part in Eq.\
(\ref{skincoulDM}) gives a non-negligible contribution to neutron
skins in effective nuclear interactions. This contribution is needed
to reproduce the neutron skin thickness values computed
self-consistently in ETF calculations of finite nuclei. We have also
seen that to leading order, both the mean location of the neutron and
proton surfaces (\ref{tDM}) and the surface width correction
(\ref{swidthns}) are basically driven by the value of the $J/Q$ ratio.
The discussions in the previous sections suggest to fit the
experimental $\Delta R_{np}^{exp}$ data by means of the following DM
inspired ansatz:
\begin{equation}
\label{skinfit}
\Delta R_{np} = 
\sqrt{\frac{3}{5}}\left( \,t - \frac{e^2 Z}{70 J} \right) 
+ \bigg( 0.3 \frac JQ + c \bigg)I ,
\end{equation}
where $t$ is given by Eq.\ (\ref{tDM}). The second term is the surface
width contribution. It is parameterized to reproduce the dashed
lines on Fig.~4, with $c=0.07$ fm or $c=-0.05$ fm.

With the ansatz (\ref{skinfit}), we will use $J/Q$ as an open
parameter. It will be constrained by a least-squares minimization from
the experimental values $\Delta R_{np}^{exp}$ derived from the
analysis of antiprotonic atoms \cite{trz01,jas04}. We note that Eq.\
(\ref{skinfit}), as well as $t$ given by Eq.\ (\ref{tDM}), depends on
the particular values of the symmetry energy at saturation $J$ and of
the nuclear matter radius $r_0$. We fix these quantities to the
empirical values $J=31.6$ MeV and $r_0=1.143$ fm (the latter
corresponds to a saturation density $\rho_0=0.16$ fm$^{-3}$). We
consider the values $c=0.07$ fm and $c=-0.05$ fm in Eq.\
(\ref{skinfit}), discussed in connection with Fig.~4, to simulate the
upper and lower bounds of the window of the theoretical predictions
for $\sigma^{sw}$ obtained according to mean-field models of nuclear
structure. In the $\chi^2$-minimization we have weighted each $\Delta
R_{np}^{exp} -\Delta R_{np}$ difference by the inverse of the
associated experimental uncertainties. That is, in practice we have
minimized the quantity
\begin{equation}
\label{chi2}
\sum_i \left(
\frac{\Delta R_{np}(i) - \Delta R_{np}^{exp}(i)}{\xi_i}
\right)^2 ,
\end{equation}
where $\Delta R_{np}$ is calculated with Eq.\ (\ref{skinfit}) and the
$\xi_i$ denote the uncertainties of the experimental data.

The fits to experiment give $J/Q = 0.667 \pm 0.047$ with $c=0.07$ fm
and $J/Q = 0.791 \pm 0.049$ with $c=-0.05$ fm (i.e., a range $0.62
\lesssim J/Q \lesssim 0.84$). The quoted uncertainties in the $J/Q$
predictions correspond to the value of one standard deviation
associated with the fit made through Eq.\ (\ref{chi2}). To check our
method of minimization and error estimation, we have applied the same
procedure to make a linear fit $m I + n$ of the experimental data
$\Delta R_{np}^{exp}$. In this case we have obtained $\Delta R_{np}=
(0.901 \pm 0.147) I + (-0.034 \pm 0.023)$ fm, which fully agrees with
the result quoted by the experimentalists \cite{trz01,jas04}.

The results for $\Delta R_{np}$ from our fits compared with the
experimental data $\Delta R_{np}^{exp}$ are displayed as a function of
$I$ in Fig.~5. Both extractions of $J/Q$, for $c=0.07$ and $c=-0.05$
fm, predict basically the same total neutron skin thickness with a
similar quality and they are close to the average $\Delta R_{np}=
(0.90 \pm 0.15) I + (-0.03 \pm 0.02)$ fm \cite{trz01,jas04} of the
experimental data. However, the splitting of the neutron skin
thickness into a part coming from the distance $t$ and another part
coming from the surface width $\Delta R_{np}^{sw}$ is different in
both cases, as we have discussed in the previous section. Therefore,
it becomes clear that the experimental neutron skin thickness data, by
themselves, may be able to constrain the total value but not its
partition into a bulk and a surface width contribution.

\begin{figure}
\includegraphics[width=0.96\columnwidth,  clip=true]{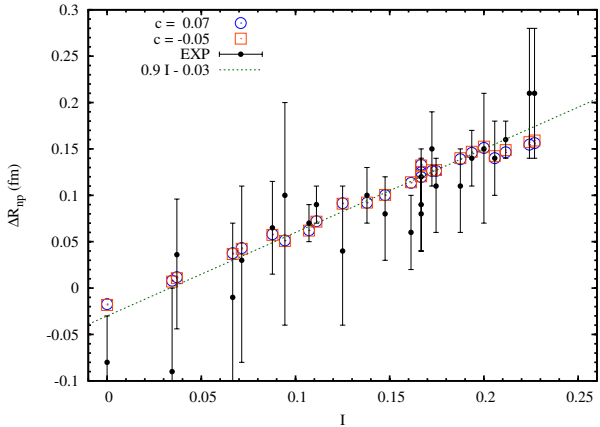}
\caption{\label{nsfig8} (Color online)  The values of $\Delta R_{np}$
obtained from Eq.\ (\ref{skinfit}) by fitting the $J/Q$ ratio, using
$c=0.07$ fm (circles) and $c=-0.05$ fm (squares), to reproduce the
$\Delta R_{np}^{exp}$ values measured in antiprotonic atoms (dots with
error-bars). The average value of $\Delta R_{np}^{exp}$ is marked by the
dotted line.}
\end{figure}

It is generally acknowledged that the value of the bulk nuclear
symmetry energy coefficient $J$ is about 30--32 MeV, but there is some
uncertainty. To assess the dependence of the extraction of $J/Q$ on
the assumed value for the $J$ coefficient (which in the above we have
taken as 31.6 MeV), we have repeated the fit of Eq.\ (\ref{skinfit})
to the neutron skin data for $J=35$ MeV and for $J=28$ MeV\@. The
results are, respectively, $J/Q = 0.642 \pm 0.046$ and $J/Q = 0.701
\pm 0.048$ if $c=0.07$ fm, and $J/Q = 0.764 \pm 0.048$ and $J/Q =
0.829 \pm 0.051$ if $c=-0.05$ fm. All in all, it seems safe to
consider that within the present model the $J/Q$ values compatible
with the data from antiprotonic atoms span a window from about
$J/Q=0.59$ to $J/Q=0.88$, or $0.6 \lesssim J/Q \lesssim 0.9$ in round
figures.

We recall that the DM of nuclei does not incorporate shell effects and
averages the corresponding quantal magnitudes. With the method
described we have fitted $\Delta R_{np}^{exp}$ that in general
contains shell effects, and possible correlation and deformation
contributions. But, as mentioned, the neutron skin data analyzed show
a well defined linear trend with the relative neutron excess $I$
(namely, $\Delta R_{np}= (0.90 \pm 0.15) I + (-0.03 \pm 0.02)$ fm
\cite{trz01,jas04}). This trend, and the agreement of the DM values
for $\Delta R_{np}$ with the self-consistent ETF calculations in
finite nuclei, which also are free of shell effects and have a
linear trend with $I$ (Figs.~2 and 3), gives more reliability to the
predictions obtained with the DM formula from the experimental data.

\subsection{Constraints on the {\boldmath$L$} parameter}

As we have discussed previously, the parameter $L$ has a direct
relation with the slope of the symmetry energy of the nuclear EOS, see
Eq.\ (\ref{slopesym}). Having determined the values of the $J/Q$ ratio
compatible with the experimental neutron skins measured in
antiprotonic atoms, we can use the linear correlation between $L$ and
$J/Q$ found in mean-field effective nuclear interactions, to obtain an
insight on the values of the parameter $L$ favored by neutron skins.

We have displayed the linear correlation $L= m J/Q + n$ in the
rightmost panel of Fig.~1 (the linear correlation coefficient for the
shown interactions is $r=0.978$). The values of the $m$ and $n$
coefficients have some dependence on the set of interactions chosen to
make the linear regression. We have checked that this dependence is
rather weak. Namely, we have tested the correlation of $L$ with $J/Q$
by taking into account successively 10, 14, 18, and 24 interactions
and we have found the linear regressions to be comprised between $L=
139 J/Q - 52$ MeV and $L= 150 J/Q - 57$ MeV\@. Considering these two
limiting cases and the constraint $0.6 \lesssim J/Q \lesssim 0.9$
found in the previous section, leads to a variation of $L$ between 31
MeV and 78 MeV\@. Thus, our estimate for the $L$ coefficient, which
takes into account the surface width correction in $\Delta R_{np}$
obtained in the calculations with mean-field interactions, basically
lies in the range $30 \lesssim L \lesssim 80$ MeV\@. Had we kept the
value of $J$ fixed at $31.6$ MeV, the extracted range for $L$ would be
a little narrower: $35 \lesssim L \lesssim 70$ MeV\@.

In a previous work \cite{cen09} we have investigated the correlations
between the symmetry energy coefficient in finite nuclei and in the
EOS at subsaturation densities. These correlations allow one to derive
an aproximate formula for the neutron skin thickness with explicit
dependence on the $L$ coefficient. By comparison of that result for
the neutron skin thickness with the experimental data set of Refs.\
\cite{trz01,jas04}, in Ref.\ \cite{cen09} we found a range of values
$L= 55\pm25$ MeV (displayed in Fig.~3 of Ref.\ \cite{cen09}) when one
includes the surface width contribution $\Delta R_{np}^{sw}$ in the
calculations. That prediction is consistent with the values obtained
here by the present procedure. A somewhat higher range $L= 75\pm25$
MeV was obtained \cite{cen09} when one neglects $\Delta R_{np}^{sw}$.
Although a vanishing $\Delta R_{np}^{sw}$ value, corresponding to
$b_n=b_p$ in the nucleon density distributions, is not favored by the
mean field interactions (see Section III), it cannot be discarded
without having more experimental evidence on the value of $b_n$ as we
have noted in Section III (see also Refs.\ \cite{trz01,swi05}).

In recent years, considerable advances in probing experimentally the
density dependence of the symmetry energy at subsaturation have been
achieved in heavy ion collisions (HIC) at intermediate energies. It
has been found that the symmetry energy can be modelized around the
saturation density with reasonable good approximation
by~\cite{tsa04,che05,li05b,ste05,lie07,li08,she07,she07a,sou06,fam06} 
\begin{equation}
c_{sym}(\rho) = J {\bigg( \frac{\rho}{\rho_0} \bigg)}^{\gamma} .
\label{esymm}
\end{equation}
From Eq.\ (\ref{esymm}), one can estimate the parameter $L$ defined in
Eq.~(\ref{l_ksym}) from the stiffness $\gamma$ of the symmetry energy,
as $L= 3\gamma J$. The values for $\gamma$ extracted in the literature
from different HIC observables fall in the range $\gamma \sim
0.55-1.05$, which implies $L$ values roughly between 50 and 100 MeV\@.
In these studies, the value of $J$ in Eq.\ (\ref{esymm}) normally has
been taken equal to 31.6 or 32 MeV\@. The extraction of the equation
of state of cold nuclear matter from HIC data is a very complicated
task and requires model assumptions
\cite{tsa04,che05,li05b,ste05,lie07,li08, 
sou06,she07,she07a,fam06,gai04,bar05,tsa09,kow07,sam07,sou08}; the
indicated estimates for $\gamma$ and $L$ may be somewhat modified as
more measurements and analyses be performed. The range $30 \lesssim L
\lesssim 80$ MeV of $L$ values determined here from neutron skins,
assuming the dependence of Eq.~(\ref{esymm}), favors a constraint
$0.32 \lesssim \gamma \lesssim 0.84$ for the $\gamma$ exponent. Thus,
the result points towards a soft symmetry energy.

Our estimates for the stiffness $\gamma$ can be compared with
alternative predictions derived in the recent literature. For
instance, our range $0.32 \lesssim \gamma \lesssim 0.84$ overlaps with
the values $\gamma\sim0.55$--0.77 that Danielewicz \cite{dan03}
obtains from the study of binding energies, neutron skins, and isospin
analog states of selected nuclei. It also contains the value $\gamma
\sim 0.55$ that is inferred from the analysis of neutron-proton
emission ratios in HIC carried out by Famiano et al.\ \cite{fam06}, as
well as with the value $\gamma \sim 0.69$ obtained by Shetty et al.\
\cite{sou06,she07,she07a} from isotopic scaling in intermediate-energy
nuclear reactions.

The stiffness of the symmetry energy at subsaturation densities also
has been investigated from isospin diffusion data in HIC, by means of
simulations with an isospin- and momentum-dependent transport model
with in-medium nucleon-nucleon cross sections
\cite{che05,ste05,li05b,lie07,li08}. In this case the prediction is
$0.69 \lesssim \gamma \lesssim 1.05$. It corresponds to a behavior of
$c_{sym}(\rho)$ that nearly ranges between the familiar $\rho^{2/3}$
dependence of the purely kinetic symmetry energy of a free Fermi gas,
in the lower limit, and linearity in the density $\rho$, in the upper
limit. The constraints on the stiffness of the symmetry energy derived
from isospin diffusion, combined with an analysis of the properties of
Skyrme interactions, are found to lead to a constraint $63 \lesssim L
\lesssim 113$ MeV \cite{che05,ste05,li05b,lie07,li08}. The predictions
from isospin diffusion are thus a little stiffer, though the lower
limits of $\gamma$ and $L$ obtained using this method are in agreement
with the upper limits of $\gamma$ and $L$ obtained from our study of
neutron skins with inclusion of the surface width contribution.

\begin{figure}
\includegraphics[width=0.80\columnwidth,angle=0,clip=true]{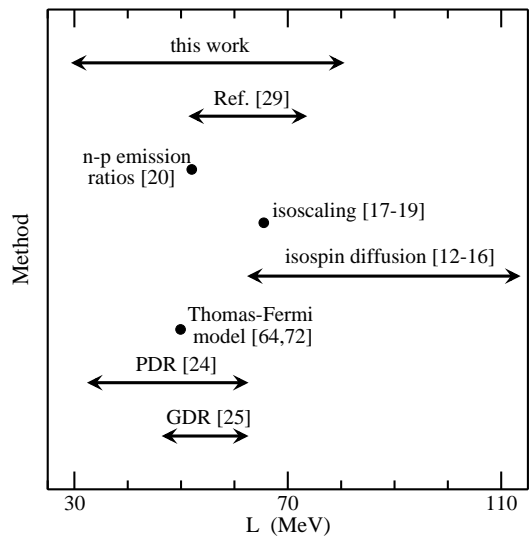}
\caption{Comparison of the estimated values of the parameter $L$ from
different observables and methods. Some of the estimates have been
analyzed through Eq.\ (\ref{esymm}) for $c_{sym}(\rho)$.}
\label{fig6}
\end{figure}

Another valuable reference comes from the celebrated Thomas-Fermi
model of Myers and \'{S}wi\c{a}tecki \cite{TF94,TF94b}. This model was
fitted very precisely to the binding energies of a comprehensive set
of 1654 nuclei. It predicts an EOS that leads to a coefficient
$L=49.9$ MeV\@. Note that if we compare $c_{sym}(\rho)$ calculated
from the EOS of the Thomas-Fermi model with Eq.\ (\ref{esymm}), an
exponent $\gamma=0.51$ is obtained. Additional information on the
density content of the symmetry energy arises from the constraints on
the symmetry pressure $P_{sym}= \rho_0 L/3$ extracted by Klimkiewicz
et al.\ \cite{kli07} from the properties of pygmy dipole resonances in
nuclei. These are indicative of a value $\gamma\sim 0.35$--0.65 if one
assumes $P_{sym}=\rho_0\gamma J$ following from Eq.\ (\ref{esymm})
given above. On the other hand, Trippa et al.\ \cite{tri08} have
obtained the constraint $23.3 < c_{sym} (\rho\!=\!0.1 \, {\rm
fm}^{-3}) < 24.9$ MeV from consideration of the giant dipole resonance
in $^{208}$Pb, which implies a range $\sim\!0.5$--0.65 for the
$\gamma$ exponent. We depict in Fig.~\ref{fig6} the estimated ranges
of values for the $L$ parameter from the discussed analyses.

In summary, in spite of the discrepancies in the details, the various
findings from experimental isospin-sensitive signals, including ours,
agree all on a rather soft nuclear symmetry energy at subsaturation
densities. Recent studies of pure neutron matter at low densities
based on universal properties of dilute Fermi gases lead to a similar
conclusion~\cite{schw05,piek07}. One may mention that there exists
recent circumstantial evidence \cite{li09}, derived from $\pi^-/\pi^+$
ratios in central HIC collisions at SIS/GSI energies, hinting at that
the nuclear symmetry energy is soft also in the regime of
suprasaturation densities ($\rho\ge 2\rho_0$). However, further
experimental and theoretical confirmations of this fact need to be
awaited \cite{li09}.

\section{Summary and conclusions}

The droplet model predicts that the neutron skin thickness of atomic
nuclei is correlated with the ratio $J/Q$, where $J$ is the symmetry
energy in bulk matter and $Q$ is the surface stiffness coefficient. We
have shown that the $J/Q$ ratio displays a linear relationship with
the DM parameter $L$ in nuclear mean field models that are calibrated
to experimental ground-state properties such as binding energies,
charge radii, and single-particle data. In this way, the known
correlation between the neutron skin thickness in a heavy nucleus and
the density derivative of the symmetry energy (or of the neutron
equation of state) evaluated at a subsaturation density, can be
interpreted in the context of the DM.

According to the droplet model, the neutron skin thickness is
correlated with the overall relative neutron excess $I=(N-Z)/A$ of
nuclei. This fact is in agreement with the experimental findings using
information from antiprotonic atoms. The DM expression for the neutron
skin thickness contains ``bulk'' and ``surface'' parts. The bulk part
corresponds to the contribution proportional to the distance $t$
between the neutron and proton mean surface locations. This part is
quite dependent on the Skyrme force or RMF parameterization used to
compute it (see Fig.~2). In finite nuclei, this DM bulk contribution
systematically underestimates the neutron skin thickness extracted
directly as the difference of the neutron and proton rms radii of the
nucleus from self-consistent ETF calculations with effective forces.
This evidence indicates that the surface part, due to the $b_n\neq
b_p$ contribution, is necessary in the DM formula to properly estimate
the neutron skin thickness in finite nuclei in mean-field models using
effective nuclear interactions.

The DM surface contribution $\Delta R_{np}^{sw}$ to the neutron skin
thickness is smaller than the bulk part and shows a well defined linear
increasing tendency with the overall relative neutron excess $I$. We
have investigated the dependence of the slope $\sigma^{sw}$ with
respect to $I$ of this surface contribution using various Skyrme and
RMF forces. We have found that the slopes $\sigma^{sw}$ lie in a
region of the \mbox{$\sigma^{sw}$--$J/Q$} plane that can be roughly
limited by two straight lines as a function of the $J/Q$ value. It
implies that nuclear properties, which are included in the calibration
of the free parameters of the Skyrme and RMF interactions, constrain
the possible values of the surface width contribution to the neutron
skin thickness in the DM\@. If the same nuclear interaction is used, a
good agreement between both the DM formula and the self-consistent ETF
calculations of $\Delta R_{np}$ in finite nuclei is found along the
whole periodic table when the contribution $\Delta R_{np}^{sw}$ is
included in the DM.

To analyze possible bounds suggested by experimental neutron skin data
on the value of the $J/Q$ ratio, we have adjusted the DM neutron skin
thickness formula to the neutron skin sizes measured in antiprotonic
atoms \cite{trz01,jas04}. We have determined a window $0.6 \lesssim
J/Q \lesssim 0.9$ for the $J/Q$ ratio by using the largest and
smallest surface contributions $\Delta R_{np}^{sw}$ obtained from
successful Skyrme forces and RMF parameterizations. These two fits
reproduce the experimental data with almost the same quality. In other
words, the experimental data of the neutron skin thickness in finite
nuclei constrain the total theoretical estimate but not its partition
into a bulk and a surface contribution. Once the window of $J/Q$
values is known, the compatible range of values of the parameter $L$
can be estimated from the linear correlation between $L$ and $J/Q$
shown in Fig.~1. From our analysis we find the constraints $30\lesssim
L\lesssim 80$ MeV.

If a model symmetry energy $c_{sym}(\rho)= J(\rho/\rho_0)^\gamma$ is
assumed, a prediction for the value of the stiffness $\gamma$ of the
symmetry energy can be obtained, with use of the empirical values of
$J$ and $\rho_0$. In this way we find the estimate
$0.32\lesssim\gamma\lesssim 0.84$. Thus, our analysis of the
experimental neutron skins deduced from antiprotonic atoms suggests a
relatively soft symmetry energy, in good accord with the recent
indications from pygmy \cite{kli07} and giant \cite{tri08} dipole
resonances. Our prediction for the stiffness $\gamma$ of the symmetry
energy also is in reasonable agreement with the constraints derived by
Danielewicz \cite{dan03}, by Famiano et al.\ \cite{fam06}, and by
Shetty et al.\ \cite{she07a} using different observables, as well as
with the value $\gamma=0.51$ of the EOS of the Thomas-Fermi model of
Myers and \'{S}wi\c{a}tecki \cite{TF94,TF94b}. In the upper limit, our
prediction overlaps with the lower limit provided by the analysis of
isospin diffusion data in intermediate-energy heavy ion collisions
\cite{che05,ste05,li05b,lie07,li08}. 

In summary, different techniques of extracting the parameters that
describe the density dependence of the symmetry energy predict different
values. However, taking the average of the central values of the
predictions displayed in Fig.~\ref{fig6}, with account of their
uncertainties when available, there exist narrow windows of the
parameters $L\sim 45$--75 MeV and $\gamma\sim 0.5$--0.8 which are,
actually, compatible with all the different methods mentioned to
obtain them. These ranges of values of the indicated parameters for
describing the leading density dependence of the symmetry energy in
bulk matter seem to be the ``optimal'' ones according to present
experimental evidence from nuclear data.

To conclude, we are aware that the neutron skin thickness data derived
from antiprotonic atoms are to some extent model dependent, and have
for some nuclei large error-bars. Also, our theoretical method
represents just an average approximation. In spite of these
limitations, we hope to have shown that from neutron skin data it is
possible to make a reasonable estimate of the density dependence of
the symmetry energy with uncertainties that are not significantly much
larger than those currently obtained from other experimental
observables. One expects that future data from the planned
parity-violating electron scattering experiment for measuring the
neutron radius in $^{208}$Pb \cite{prex} will contribute to narrow
down the constraints derived here from the thickness of neutron skins
of nuclei.

\begin{acknowledgments}
The authors are very grateful to W. J. \'{S}wi\c{a}tecki for valuable
discussions during the XIV Nuclear Physics Workshop held in
Kazimierz-Dolny (Poland). Helpful correspondence with A. Trzci{\'n}ska
is gratefully acknowledged. Work partially supported by Grants No.\
FIS2005-03142 and FIS2008-01661 from MEC (Spain) and FEDER, No.\
2009SGR-1289 from Generalitat de Catalunya, and by the Spanish
Consolider-Ingenio 2010 Programme CPAN CSD2007-00042. X.R. also
supported by grant AP2005-4751 from MEC (Spain). M.W. gratefully
acknowledges the warm hospitality extended to him by the Nuclear
Theory Group and the Departament d'Estructura i Constituents de la
Mat\`eria at the University of Barcelona during his stay in Barcelona.
\end{acknowledgments}

\section{Appendix}

To compute the surface stiffness coefficient $Q$, and also the neutron
and proton surface widths $b_n$ and $b_p$ that appear in the
contribution $\Delta R_{np}^{sw}$ of Eq.\ (\ref{swidthns}) to the
neutron skin thickness, we obtain the self-consistent neutron and
proton density profiles in asymmetric semi-infinite nuclear matter
(ASINM). To do that we consider a semi-infinite slab with a plane
interface separating a mixture of protons and neutrons at the left,
whose densities decrease smoothly to zero at the right as empty space
is reached. The axis perpendicular to the interface is taken to be the
$z$ axis. Thus, the relative neutron excess $\delta= (\rho_n-\rho_p)/
(\rho_n+\rho_p)$ depends locally on the $z$ coordinate. When $z$ goes
to minus infinity, the neutron and proton densities approach the
values of asymmetric uniform nuclear matter in equilibrium,
corresponding to an interior bulk neutron excess~$\delta_0$.

In order to obtain the proton and neutron densities in ASINM one has
to minimize the total energy per unit area with respect to arbitrary
variations of the densities, with the constraint of conservation of
the number of protons and neutrons. When $\delta_0$ is not very large,
so that occurrence of drip nucleons does not take place (which is the
situation in all cases considered in the present work), the
constrained energy per unit area reads \cite{mye85,kol85,cen98a}
\begin{equation}
\frac{E}{S} = \int^{\infty}_{-\infty} 
\big[ \varepsilon(z) - \mu_n \rho_n(z) - \mu_p \rho_p(z) \big] dz .
\label{profile}
\end{equation}
In this equation, $\varepsilon(z)$ represents the energy density
functional of the nuclear effective interaction under investigation,
and $\mu_n$ and $\mu_p$ are the neutron and proton chemical
potentials. The explicit expressions of $\varepsilon(z)$ in the
extended Thomas-Fermi (ETF) method for Skyrme forces and relativistic
mean field interactions can be found in e.g.\ Appendix~A of Ref.\
\cite{cen98a}. The proton and neutron densities obey the coupled local
Euler-Lagrange equations
\begin{equation}
\frac{\delta \varepsilon(z)}{\delta \rho_n} - \mu_n = 0, \qquad
\frac{\delta \varepsilon(z)}{\delta \rho_p} - \mu_p = 0.
\label{eulernr}
\end{equation}
We solve them fully self-consistently by numerical iteration
\cite{cen90}. We note in this respect that we have not used any
parameterized form of the densities such as e.g.\ Fermi shapes. In
the relativistic model the variational equations (\ref{eulernr}) are
supplemented with additional field equations for the meson fields; the
calculational details for the relativistic problem can be found in
Refs.\ \cite{cen90,cen93,cen98a}.

From the calculated density profiles, one obtains the mean locations
of the surfaces, $z_{0q}$ ($q=n,p$), as
\begin{equation}
z_{oq} = \frac{\int^{\infty}_{- \infty} z \rho'_q(z)dz}
{\int^{\infty}_{- \infty} \rho'_q(z)dz},
\label{locations}
\end{equation}
where the primes indicate a derivative with respect to the $z$ coordinate.
The proton and neutron surface widths in ASINM are obtained as the second
moment of $\rho'(z)$:
\begin{equation}
b^2_q = \frac{\int^{\infty}_{- \infty} (z - z_{0q})^2 \rho'_q(z)dz}
{\int^{\infty}_{- \infty} \rho'_q(z)dz}.
\label{widths}
\end{equation}

The distance $t= z_{0n} - z_{0p}$ between the mean surface locations
of the neutron and proton density profiles allows one to
extract the surface stiffness coefficient $Q$ from the relation
\cite{mye69} 
\begin{equation}
t= z_{0n} - z_{0p} = \frac{3 r_0}{2}  \frac{J}{Q}  \delta_0 ,
\label{sthickness}
\end{equation}
which is valid in the limit of small asymmetries. For each given
nuclear interaction we solve the ASINM problem for 5 different values
of $\delta_0$, between 0.005 and 0.025, and we then evaluate $Q$ from
the slope of $t$. 

There exists a second way of computing $Q$. It is based on the fact
that in dividing the energy in bulk and surface parts, as soon as
$\delta_0 \neq 0$, there are two possibilities to define the bulk
reference energy \cite{mye85,cen98a}. One definition is based on the
chemical potentials $\mu_n$ and $\mu_p$ of each nucleon species; this
definition is thermodynamically consistent and it is the one that we
have given in Eq.\ (\ref{profile}). The second definition of the
reference energy is based on taking the value of the energy per
particle in bulk asymmetric nuclear matter. Accordingly, upon the bulk
reference energy chosen, there exist two forms of the surface energy
in ASINM, which are called $E_{\rm surf,\mu}$ and $E_{\rm surf,e}$
\cite{mye85,cen98a}. In the small asymmetry limit, it can be shown
that the difference between these two quantities behaves as
\begin{equation}
E_{\rm surf,e} - E_{\rm surf,\mu} = \frac{9 J^2}{2 Q} \delta_0^2 .
\label{esurfemu}
\end{equation}
Thus, the slope of $E_{\rm surf,e} - E_{\rm surf,\mu}$ with respect to
$\delta_0^2$ provides another means to extract the value of the
surface stiffness coefficient $Q$. We have computed $Q$ from Eq.\
(\ref{sthickness}) and have used Eq.\ (\ref{esurfemu}) to confirm the
validity of our calculated values.



\begin{thebibliography}{99}


\bibitem{ang04} I. Angeli, 
                At.\ Data Nucl.\ Data Tables {\bf 87}, 185 (2004).
\bibitem{prex}  R. Michaels, P.~A. Souder, and G.~M. Urciuoli, spokespersons,
                Jefferson Laboratory Experiment E06002,
                http://hallaweb.jlab.org/parity/prex.
\bibitem{lat04} J.~M. Lattimer and M. Prakash,
                 Science {\bf 304}, 536 (2004);
                Phys. Rep. {\bf 442}, 109 (2007).
\bibitem{hei00} H. Heiselberg and M. Hjorth-Jensen,
                Phys.\ Rep.\ {\bf 328}, 237 (2000).
\bibitem{ste05a}A.~W. Steiner, M. Prakash, J.~M. Lattimer and P.~J. Ellis,
                Phys. Rep. {\bf 411}, 325 (2005).
%
\bibitem{sil05} T. Sil, M. Centelles,  X. Vi\~nas, and J. Piekarewicz,
		Phys. Rev. {\bf C71}, 045502 (2005), and references therein.
\bibitem{gravi} P. Jofr\'e, A. Reisenegger, and R. Fern\'andez,
                Phys. Rev. Lett. {\bf 97}, 131102 (2006);
                P.G. Krastev and Bao-An Li,
                Phys. Rev. {\bf C 76}, 055804 (2007).
%
\bibitem{khoa96} Dao T. Khoa, W. von Oertzen, and A. A. Ogloblin,
                 Nucl.\ Phys.\ {\bf A602}, 98 (1996).
\bibitem{gai04} T. Gaitanos, M. Di Toro, S. Typel, V. Baran,
                C. Fuchs, V. Greco, and H. H. Wolter, 
                Nucl. Phys. {\bf A732}, 24 (2004).
%
\bibitem{bar05} V. Baran, M. Colonna, V. Greco, and M. Di Toro, 
                Phys.\ Rep.\ {\bf 410}, 335 (2005).
%
\bibitem{tsa04} M. B. Tsang et al., 
                Phys. Rev. Lett.  {\bf 92}, 062701 (2004).
%
\bibitem{ste05} A.~W. Steiner and B. A. Li,
                Phys. Rev. {\bf C72}, 041601(R) (2005).
\bibitem{che05} L. W. Chen, C. M. Ko, and B. A. Li,
                Phys. Rev. Lett. {\bf 94}, 032701 (2005).
\bibitem{li05b} L. W. Chen, C. M. Ko, and B. A. Li,
                Phys.\ Rev.\ {\bf C72}, 064309 (2005).
\bibitem{lie07} L. W. Chen, C. M. Ko, and B. A. Li,
                Phys.\ Rev.\ {\bf C76}, 054316 (2007).
%
\bibitem{li08}  B. A. Li, L. W. Chen, and C. M. Ko,
                Phys.\ Rep.\ {\bf 464}, 113 (2008).
\bibitem{sou06} G. A. Souliotis, D. V. Shetty, A. Keksis, E. Bell, 
                M. Jandel, M. Veselsky and S. J. Yennello,
                Phys. Rev. {\bf C73}, 024606 (2006).
\bibitem{she07} D.~V. Shetty, S.~J. Yennello, G.~A. Souliotis,
                Phys.\ Rev.\ {\bf C75}, 034602 (2007).
\bibitem{she07a} D.~V. Shetty, S.~J. Yennello, G.~A. Souliotis,
                Phys.\ Rev.\ {\bf C76}, 024606 (2007).             
\bibitem{fam06} M. A. Famiano et al., 
                Phys. Rev. Lett. {\bf 97}, 052701 (2006). 
%
\bibitem{tsa09} M. B. Tsang et al.,
                Phys. Rev. Lett.  {\bf 102}, 122701 (2009).
%
\bibitem{pie02} J. Piekarewicz, 
                Phys. Rev. {\bf C66}, 034305 (2002);
                Phys. Rev. {\bf C69}, 041301(R) (2004).
\bibitem{li07}  T. Li et al.,
                Phys.\ Rev.\ Lett.\ {\bf 99}, 162503 (2007).
%
\bibitem{kli07} A. Klimkiewicz et al,
                Phys.\ Rev.\ {\bf C76}, 051603(R) (2007).
%
\bibitem{tri08} L. Trippa, G. Col\`o, and E. Vigezzi,
                Phys.\ Rev.\ {\bf C77}, 061304(R) (2008).
%
\bibitem{lia08} H. Liang, N. Van Giai, and J. Meng,
                Phys.\ Rev.\ Lett.\ {\bf 101}, 122502 (2008).
%
\bibitem{mye80} W. D. Myers and W. J. \'{S}wi\c{a}tecki, 
                Nucl. Phys. {\bf A336}, 267 (1980).
%
\bibitem{swi05} W.~J. \'{S}wi\c{a}tecki, A. Trzci{\'n}ska and J. Jastrz\c{e}bski,
                Phys. Rev. {\bf C71}, 047301 (2005).
\bibitem{dan03}	P. Danielewicz,
		Nucl. Phys. {\bf A727}, 233 (2003).
%
\bibitem{dan09} P. Danielewicz and J. Lee,
                Nucl. Phys. {\bf A818}, 36 (2009).
\bibitem{pet96} C.~J. Pethick and D.~G. Ravenhall,
                Nucl. Phys. {\bf A606}, 173 (1996).
\bibitem{miz00} S. Mizutori, J. Dobaczewski, G. A. Lalazissis, 
		W. Nazarewicz and P.-G. Reinhard,
		Phys. Rev. {\bf C61}, 044326 (2000).
\bibitem{fur02} R. J. Furnstahl,
               	Nucl. Phys. {\bf A706}, 85 (2002).
\bibitem{tre86}	J. Treiner and H. Krivine,
		Ann. of Phys. {\bf 170}, 406 (1986).
\bibitem{war98} M. Warda, B. Nerlo-Pomorska and K. Pomorski,
                Nucl. Phys. {\bf A635}, 484 (1998).
\bibitem{die03}	A. E. L. Dieperink, Y. Dewulf, D. Van Neck, 
		M. Waroquier, and V. Rodin,
		Phys. Rev. {\bf C68}, 064307 (2003).
\bibitem{bro00} B.~A. Brown, 
                Phys. Rev. Lett. {\bf 85}, 5296 (2000).
\bibitem{typ01} S. Typel and B.~A. Brown, 
                Phys. Rev. {\bf C64}, 027302 (2001).
\bibitem{cen02} M. Centelles, M. Del Estal, X. Vi\~nas, and S.~K. Patra,
                in \textit{The Nuclear Many-Body Problem {\sl 2001}},
                Vol.\ 53 of NATO Advanced Studies Institute Series B:
                Physics, edited by W. Nazarewicz and D. Vretenar
                (Kluwer, Dordrecht, 2002), p.\ 97. 
\bibitem{bal04} M. Baldo, C. Maieron, P. Schuck and X. Vi\~nas, 
                Nucl. Phys. {\bf A736}, 241 (2004).
\bibitem{ava07} S. S. Avancini, J. R. Marinelli, D. P. Menezes, 
		M. M. W. Moraes, and C. Provid\^{e}ncia,
 		Phys. Rev. {\bf C75}, 055805 (2007).
\bibitem{ray78} L. Ray, G. W. Hoffmann, G.~S. Blanpied , W.~R. Coker, and R.~P. Liljestrand,
                Phys. Rev. {\bf C18}, 1756 (1978);
                L. Ray and G.~W. Hoffmann,
                Phys. Rev. {\bf C31}, 538 (1985).
\bibitem{sta94}
                V.~E. Starodubsky
                and  N.~M. Hintz,
                Phys.\ Rev.\  \textbf{C49}, 2118 (1994).                
\bibitem{kar02} S. Karataglidis, K. Amos, B. A. Brown and P. K. Deb,            
                Phys. Rev. {\bf C65}, 044306 (2002).
                
\bibitem{cla03} B. C. Clark, L. J. Kerr and    S. Hama, 
                Phys. Rev. {\bf C67},  054605 (2003).                
\bibitem{kra99} A. Krasznahorkay et al., 
                Phys. Rev. Lett. {\bf 82}, 3216 (1999).
\bibitem{kra04}	A. Krasznahorkay et al., 
		Nucl. Phys. {\bf A731}, 224 (2004).
\bibitem{trz01}
                A. Trzci{\'n}ska,
                J. Jastrz\c{e}bski,
                P. Lubi{\'n}ski,
                F.~J. Hartmann,
                R. Schmidt,
                T. von Egidy
                and  B. K{\l}os,
                Phys.\ Rev.\ Lett. \textbf{87}, 082501 (2001).
\bibitem{jas04} J. Jastrz\c{e}bski,
                A. Trzci{\'n}ska,
                P. Lubi{\'n}ski,
                B. K{\l}os,
                F.~J. Hartmann,
                T. von Egidy and
                S. Wycech,
		Int. J. Mod. Phys. {\bf E13}, 343 (2004).
\bibitem{klo07}
                B. K{\l}os et al.,
                Phys.\ Rev.\  \textbf{C76}, 014311 (2007).
\bibitem{cen09} M. Centelles, X. Roca-Maza, X. Vi\~nas, and M. Warda,
                Phys.\ Rev.\ Lett.\ {\bf 102}, 122502 (2009).
\bibitem{mye69} W.~D. Myers and W.~J. \'{S}wi\c{a}tecki, 
                Ann. of Phys. (N.Y.) {\bf 55}, 395 (1969);
                Ann. of Phys. (N.Y.) {\bf 84}, 186 (1974).
%
\bibitem{mye77} W.~D. Myers, ``Droplet Model of Atomic Nuclei''
                (Plenum, New York, 1977).
\bibitem{bra85} M. Brack, C. Guet and H.-B. H\aa kansson, 
		Phys. Rep. {\bf 123}, 275 (1985).
\bibitem{far78}	M. Farine, J. M. Pearson and B. Rouben,
 		Nucl. Phys. {\bf A304}, 317 (1978);
%
		M. Farine, J. C\^{o}t\'{e} and J. M. Pearson
 		Nucl. Phys. {\bf A338}, 86 (1980);
		Phys. Rev. {\bf C24}, 303 (1981).
\bibitem{kol85} K. Kolehmainen, M. Prakash, J. M. Lattimer, and J. Treiner,
                Nucl. Phys. {\bf 439}, 535 (1985).
\bibitem{cen93a}M. Centelles and X. Vi\~nas,
                Nucl. Phys. {\bf A563}, 173 (1993).
\bibitem{cen98a}M. Centelles, M. Del Estal and X. Vi\~nas, 
                Nucl. Phys. {\bf A635}, 193 (1998).
%
\bibitem{piek09} J. Piekarewicz and M. Centelles,
                 Phys.\ Rev.\ {\bf C79}, 054311 (2009).
\bibitem{lop88} M. Lopez-Quelle, S. Marcos, R. Niembro, A. Bouyssy, 
		Nguyen Van Giai,
                Nucl.\ Phys.\ {\bf A483}, 479 (1988).
%
\bibitem{bednarek09} I. Bednarek and R. Manka, arXiv:0905.0131.
\bibitem{cen90}	M. Centelles, M. Pi, X. Vi\~nas, F. Garcias and M. Barranco,
                Nucl. Phys. {\bf A510}, 397 (1990).
\bibitem{cen93}	M. Centelles, X. Vi\~nas, M. Barranco and P. Schuck,
                Ann. of Phys. (N.Y.) {\bf 221}, 165 (1993).
\bibitem{TF94} W. D. Myers and W. J. \'{S}wi\c{a}tecki,
                Nucl. Phys. {\bf A601}, 141 (1996).
\bibitem{bla80} J. P. Blaizot, 
		Phys.\ Rep.\ {\bf 64}, 171 (1980).
%
\bibitem{patra02b}
S. K. Patra, M. Centelles, X. Vi\~nas, and M. Del Estal,
Phys.\ Rev.\ {\bf C65}, 044304 (2002).
%
\bibitem{colo04} G. Col\`o, N. Van Giai, J. Meyer, K. Bennaceur,
and P. Bonche, Phys.\ Rev.\ {\bf C70}, 024307 (2004).
\bibitem{mye85} W. D. Myers, W. J. \'{S}wi\c{a}tecki, and C. S. Wong,
                Nucl. Phys. {\bf A436}, 185 (1985).
\bibitem{kow07} S. Kowalski et al., 
                Phys.\ Rev.\ {\bf C75}, 014601 (2007).
\bibitem{sam07} S. K. Samaddar, J. N. De, X. Vi\~nas, and M. Centelles,
                Phys.\ Rev.\ {\bf C76}, 041602(R) (2007);
                Phys.\ Rev.\ {\bf C78}, 034607, (2008).
\bibitem{sou08} S. R. Souza, M. B. Tsang, R. Donangelo, W. G. Lynch, 
                and A. W. Steiner,
                Phys.\ Rev.\ {\bf C78}, 014605 (2008).
%
\bibitem{TF94b}  W. D. Myers and W. J. \'{S}wi\c{a}tecki,
                 Phys.\ Rev.\ {\bf C57}, 3020 (1998).
%
\bibitem{schw05} A. Schwenk and C. J. Pethick,
                 Phys.\ Rev.\ Lett.\ {\bf 95}, 160401 (2005).
%
\bibitem{piek07} J. Piekarewicz, 
                 Phys.\ Rev.\ {\bf C76}, 064310 (2007).
%
\bibitem{li09} Z. Xiao, B. A. Li, L. W. Chen, G. C. Yong, and M. Zhang,
               Phys.\ Rev.\ Lett.\ {\bf 102}, 062502 (2009).
%
\end{thebibliography}

\end{document}